%&amstex
%Paper: solv-int/9511006
%From: "Ravil I. Yamilov" <yamilov@nkc.bashkiria.su>
%Date: Thu, 16 Nov 95 14:58:04 +0500

%-------------------------------- This is AmsTeX file --------------
\NoBlackBoxes
\pagewidth{13.5cm}
\pageheight{17cm}
%------------------------------
\def\pa{\partial}
\def\ti{\tilde}

\def\res{\hbox{res}}
%------------------------------
\def\al{\alpha}
\def\be{\beta}
\def\ga{\gamma}
\def\de{\delta}
\def\De{\Delta}
\def\ep{\varepsilon}

\def\la{\lambda}

\def\ph{\varphi}
\def\Ph{\Phi}
\def\om{\omega}

%%-------------------------------------- Title ----------------------
\centerline{\bf CLASSIFICATION OF EVOLUTIONARY EQUATIONS ON THE LATTICE}
\centerline{\bf I. THE GENERAL THEORY}
\bigskip
\medskip
\centerline{\bf D. Levi and R. Yamilov}
\bigskip
\noindent
Dipartimento di Fisica -- Terza Universit\'a di Roma and INFN-Sezione
di Roma, P-le A. Moro 2, 00185 Roma, Italy. E-mail: levi\@roma1.infn.it
\medskip
\noindent
Ufa Institute of Mathematics, Russian Academy of Sciences,
Chernyshevsky str. 112, Ufa 450000, Russia.
E-mail: yamilov\@nkc.bashkiria.su
%%-------------------------------------- Introduction ---------------
\bigskip
\medskip
\par
{\bf 1. Introduction}
\medskip
\par
Nonlinear differential-difference equations are always more important
in applications. They enter as models for many biological chains, are
encountered frequently in queuing-problems and as discretisations of
field theories. So, both as themselves and as
approximations of continuous problems,
they play a very important role in many fields of mathematics, physics,
biology and engineering. Not many tools are available to solve such
kind of problems. Apart for a few exceptional cases the solution of
nonlinear differential-difference equations can be obtained only by
numerical calculations or by going to the continuous limit when the
lattice spacing vanishes and the system is approximated by a continuous
nonlinear partial differential equation. Exceptional cases are those
equations which, in a way or another, are either linearisable or
integrable via the solution of an associated spectral problem on the
lattice. In all such cases we can write down a
denumerable set of exact solutions corresponding to symmetries of the
nonlinear differential-difference equations. Such symmetries can be
either depending just on the dependent field and independent variable
and are denoted as point symmetries or can depend on the dependent
field in various positions of the lattice and in this case we speak
of generalised symmetries. Any differential-difference equation can
have point symmetries, but the existence of generalised symmetries
is usually associated only to the integrable ones.
%--------------------------------------------------------------------
\par
Few classes of integrable nonlinear differential-difference equations
are known [1-4] and are important for all kind of applications both
as themselves and as starting point for perturbation analysis.
However not all cases of physical interest are covered and so it
would be nice to make the classification of integrable nonlinear
differential-difference equations to provide a basic list of equations
to be used as model of nonlinear systems on the lattice or as starting
point of perturbation theory. A way to accomplish such a goal is
provided by the so-called formal symmetry approach introduced by
A.B. Shabat and collaborators in Ufa
(see e.g. review articles [5-7]) which consists in
classifying all equations of a certain class which possess few
generalised symmetries. Such an approach has been introduced at first
to classify partial differential equations but then the procedure
has been extended to the case of differential-difference
equations [2-4].
%--------------------------------------------------------------------
\par
In such an approach, one provides conditions under which one can prove
the existence of at least one (or more) generalised symmetries. These
conditions are basic tools to start the procedure of classification,
i.e. we look for the form of the nonlinear differential-difference
equations which is compatible with these conditions. This proces give
rise to classes of equations. The list of all nonlinear integrable
differential-difference equations is in principle very long, as any
point transformation transforms an integrable equation into another
one of the same kind. Thus we will consider two equations different
only if they cannot be related by a point transformation. The class
of nonlinear differential-difference equations we will study in the
following is given by
$$
u_{n,t}(t) = f_n(u_{n-1}(t), \ u_n(t), \ u_{n+1}(t)),           \eqno(1.1)
$$
where $u_n(t)$ is a complex dependent field expressed in terms of its
dependent variables, $t$ varying over the complex numbers while $n$
varying over the integers. Eq.(1.1) is a differential functional
relation which correlates the time evolution of a function calculated
at the point $n$ to its values in its nearest neighbor points
$(n+1, \ n-1)$. A peculiarity of the choice of eq.(1.1) is the fact
that the right hand side of it is not just a function, i.e. it is not
the same for all points in the lattice, but for each point of the
lattice one has an a priori different right hand side. In such a way,
for example,
by imposing periodicity conditions on the dependent field in the
lattice variables one is able to rewrite eq.(1.1) as a coupled system
of nonlinear differential-difference equations. For example if
$$
f_{2n} = u_{2n} (u_{2n+1} - u_{2n-1}), \qquad
f_{2n+1} = u_{2n+2} - u_{2n},                                   \eqno(1.2)
$$
we can define
$$
a_n = u_{2n}, \qquad b_n = u_{2n+1},                            \eqno(1.3)
$$
and eq.(1.1) reduces to
$$
a_{n,t} = a_n (b_n - b_{n-1}), \qquad
b_{n,t} = a_{n+1} - a_n                                         \eqno(1.4)
$$
which by introducing
$$
a_n = e^{\chi_n - \chi_{n-1}}, \qquad b_n = \chi_{n,t}
$$
reduces to the Toda lattice equation
$$
\chi_{n,tt} = e^{\chi_{n+1} - \chi_n} - e^{\chi_n - \chi_{n-1}}. \eqno(1.5)
$$
%--------------------------------------------------------------------
\par
Before, in the framework of the formal symmetry approach,
only $n$-independent differen\-tial-difference equations were
considered; the following
classes of equations were completely classified:
$$
u_{n,t} = f(u_{n-1}, \ u_n, \ u_{n+1})                          \eqno(1.6)
$$
(Volterra type equations, see [2]) and
$$
u_{n,tt} = f(u_{n,t}, \ u_{n-1}, \ u_n, \ u_{n+1})             \eqno(1.7)
$$
(Toda type equations, see [3]).

If in the case of the class of equations (1.6) an equation
is defined by a function $f$, in the case of (1.1) we have an
infinite set $\{ f_n \}$ of a priori quite different functions.
So, this paper will be a next step in the classification and in the
general theory of the formal symmetry approach (readers can find
elements of a previous version of the general theory only in [3, 5]).
%--------------------------------------------------------------------
\par
Section 2 is devoted to the construction of a certain number of
conditions necessary to prove that an equation of the class (1.1)
has generalised symmetries and higher order conservation laws.
The obtained conditions are applied in Section 3 to define two
main classes of nonlinear equations possessing generalized
symmetries. The complete classification of this classes of equations
is left to a future work. Here one just shows that the two classes
are not empty as in one case we have the Toda lattice while in the
other the discrete version of the Krichever-Novikov equation.
%%-------------------------------------- Section 2 ------------------
\bigskip
\medskip
\par
{\bf 2. Construction of the classifying conditions}
\medskip
\par
If equation (1.1) is to represent a nonlinear evolutionary difference
equation, then it must depend in an essential way from the points
$(n \pm 1)$, the nearest neibouring points with respect to the point
$n$ in which we compute the time evolution. This implies that we must
add to eq.(1.1) the condition
$$
{\pa f_n \over \pa u_{n+1}} \ne 0, \qquad
{\pa f_n \over \pa u_{n-1}} \ne 0 \qquad
\hbox{for any} \ n.                                             \eqno(2.1)
$$
Before considering in detail the problem of costructing generalized
symmetries to eq.(1.1) we will introduce few definitions necessary
for the future calculations.
%--------------------------------------------------------------------
\par
A function $g_n$ depending on the set of fields $u_n$, for $n$
varying on the lattice, will be called a {\it rectricted function} and
will be denoted by the symbol RF if it is defined on a compact support,
i.e. if
$$
g_n = g_n(u_{n+i}, \ u_{n+i-1}, \ \dots, \ u_{n+j+1}, \ u_{n+j}),
\qquad i \ge j,                                                 \eqno(2.2)
$$
and $i$ and $j$ are finite integer numbers.
%--------------------------------------------------------------------
\par
If there exist, in the range
of the possible values of $n$, values $k$ and $m$ such that
$$
{\pa g_k \over \pa u_{k+i}} \ne 0, \qquad
{\pa g_m \over \pa u_{m+j}} \ne 0,                              \eqno(2.3)
$$
then we say that the function $g_n$ has a {\it length} $i-j+1$. For example,
$g_n$ could be given by the function:
$$
g_n = n u_{n+1} + u_n + [1+(-1)^n] u_{n-1};
$$
then $i = 1, \ j = -1$ and the length of $g_n$ is equal to 3 even if
only the even functions are depending on $u_{n-1}$.
%--------------------------------------------------------------------
\par
Let us define the shift operator $D$ such that
$$
D g_n(u_{n+i}, \ \dots, \ u_{n+j}) =
g_{n+1}(u_{n+i+1}, \ \dots, \ u_{n+j+1}).
$$
Then we can split the RF into equivalent classes.
%--------------------------------------------------------------------
\medskip
\noindent
{\it Definition.} Two RF
$$
a_n(u_{n+i_a}, \ \dots, \ u_{n+j_a}) \quad \hbox{and} \quad
b_n(u_{n+i_b}, \ \dots, \ u_{n+j_b})
$$
are {\it equivalent}
$$
a_n \sim b_n
$$
iff
$$
a_n - b_n = (D-1) c_n,                                          \eqno(2.4)
$$
where $c_n$ is a RF.
%--------------------------------------------------------------------
\medskip
\noindent
If, for example, we have $a_n = u_n + u_{n+1}$, it is immediate to see
that $a_n$ is equivalent to a function $b_n = 2 u_n$, as
$$
a_n - b_n = u_{n+1} - u_n = (D-1) u_n.
$$
%--------------------------------------------------------------------
\par
Let us notice that any function which is equal to a total difference
is equivalent to zero, i.e. $a_n = (D-1) c_n \sim 0$. If a RF $a_n$
of length $i-j+1$ ($i > j$) is equivalent to zero, then there will exist,
by necessity, a RF $c_n$ of the length $i-j$ such that $a_n = (D-1) c_n$.
More precisely,
$$
a_n(u_{n+i}, \dots, u_{n+j}) = c_{n+1}(u_{n+i}, \dots, u_{n+j+1}) -
c_n(u_{n+i-1}, \dots, u_{n+j}).
$$
We see that
$$
{\pa a_n \over \pa u_{n+i}} =
{\pa c_{n+1} \over \pa u_{n+i}}(u_{n+i}, \ \dots, \ u_{n+j+1}),
$$
and consequently
$$
{\pa^2 a_n \over \pa u_{n+i} \pa u_{n+j}} = 0.                  \eqno(2.5)
$$
%--------------------------------------------------------------------
\par
In the case $i = j$,
$$
a_n(u_{n+i}) = c_{n+1}(u_{n+k_1+1}, \dots, u_{n+k_2+1}) -
c_n(u_{n+k_1}, \dots, u_{n+k_2}).
$$
But as
${\pa a_n \over \pa u_{n+k_2}} = - {\pa c_n \over \pa u_{n+k_2}} = 0$
for $k_2 < i$, then $c_n = d_n(u_{n+k_1}, \dots, u_{n+i})$ and consequently
$$
a_n(u_{n+i}) = d_{n+1}(u_{n+k_1+1}, \dots, u_{n+i+1}) -
d_n(u_{n+k_1}, \dots, u_{n+i}).
$$
But also ${\pa a_n \over \pa u_{n+k_1+1}} = 0$ for $k_1 \ge i$ so that
$d_n$ cannot depend on $u_{n+k}$ for any $k$ and consequently
$$
{d a_n \over d u_{n+i}} = 0.                                    \eqno(2.6)
$$
So $a_n$ is {\it invariant} as, by definition, invariant functions are
those which depend only on $n$ and not on $u_n$.
%--------------------------------------------------------------------
\par
We can moreover define the {\it "formal" variational derivative} of a RF
$a_n$of length $i-j+1$ as:
$$
{\de a_n \over \de u_n} =
\sum_{k = n-i}^{n-j} {\pa a_k \over \pa u_n}.                   \eqno(2.7)
$$
This quantity is strictly related to the notion of variational derivative,
and this is the reason for its name. If $a_n$ is linear, then
${\de a_n \over \de u_n}$ is an invariant function, but if it is
nonlinear, then
$$
{\de a_n \over \de u_n} =
\ti g_n(u_{n+N}, \ \dots, \ u_{n-N}),                           \eqno(2.8)
$$
where for some $k$ \quad ${\pa \ti g_k \over \pa u_{k+N}} \ne 0$ and for some
$m$ \quad ${\pa \ti g_m \over \pa u_{m-N}} \ne 0$. It is immediate to prove
that if $a_n$ is a RF which is equivalent to zero, i.e. it is total
difference, i.e. it is the difference of an RF calculated in two
neighboring points of the lattice, then the formal variational derivative
of $a_n$ is zero. One can also prove that if
${\de a_n \over \de u_n} = 0$, then $a_n$ is equivalent to zero.
In fact, using eq.(2.7) we have
$$
{\pa^2 a_n \over \pa u_{n+i} \pa u_{n+j}} = 0
$$
which implies that
$$
a_n = b_n(u_{n+i}, \ \dots, \ u_{n+j+1}) +
c_n(u_{n+i-1}, \ \dots, \ u_{n+j}) \sim
d_n(u_{n+i-1}, \ \dots, \ u_{n+j}),
$$
i.e. $a_n$ is equivalent to a RF of length $i-j$. Carrying out recursively
this reasoning, we arrive at the conclusion that
$$
a_n \sim F_n(u_n) \quad \hbox{with} \quad F'_n = 0,
$$
i.e. $a_n$ must be equivalent to an $n$-dependent constant (or invariant
function) which is always equivalent to zero.
%--------------------------------------------------------------------
\par
Given a nonlinear chain (1.1) we will say that the RF
$g_n(u_{n+i}, \ \dots, \ u_{n+j})$ is a generalized (or higher) local
symmetry of our equation of {\it order} \ $i$
(more precisely, of left order $i$) iff
$$
u_{n,\tau} = g_n(u_{n+i}, \ \dots, \ u_{n+j})                  \eqno(2.9a)
$$
is compatible with (1.1), i.e. iff
$$
\pa_t \pa_{\tau} (u_n) = \pa_{\tau} \pa_t (u_n).                \eqno(2.9b)
$$
Explicitating this condition we get:
$$
\pa_t g_n = \pa_{\tau} f_n = {\pa f_n \over \pa u_{n+1}} u_{n+1, \tau}
+ {\pa f_n \over \pa u_n} u_{n, \tau} +
{\pa f_n \over \pa u_{n-1}} u_{n-1, \tau}
$$ $$
= \left[ {\pa f_n \over \pa u_{n+1}} D + {\pa f_n \over \pa u_n}
+ {\pa f_n \over \pa u_{n-1}} D^{-1} \right] g_n = f_n^* g_n,
$$ i.e. $$
(\pa_t - f_n^*) g_n = 0,                                        \eqno(2.10)
$$
where by $f_n^*$ we mean the Frechet derivative of the function $f_n$,
which is given by:
$$
f_n^* = {\pa f_n \over \pa u_{n+1}} D + {\pa f_n \over \pa u_n}
+ {\pa f_n \over \pa u_{n-1}} D^{-1} =
f_n^{(1)} D + f_n^{(0)} + f_n^{(-1)} D^{-1}.                    \eqno(2.11)
$$
Eq.(2.10) is an equation for $g_n$ once the function $f_n$ is given,
an equation for the symmetries. In this work we limit ourselves to
local symmetries, i.e. symmetries which are RF. A nonlocal extention can
be carried out by introducing, for example, a new field
$v_n$: $(D-1) v_n = u_n$, i.e. $v_n = - \sum_{j=n}^{\infty} u_j$ or
$v_n = \sum_{j=-\infty}^{n-1} u_j$ (compare [5]).
Extension in such a direction will be carried out in future work.
%--------------------------------------------------------------------
\par
Given a symmetry we can construct a new symmetry by applying a
recursive operator, i.e. an operator which transforms a symmetry
into another one. An operator
$$
L_n = \sum_{j = -\infty}^{m} l_n^{(j)}(t) D^j                   \eqno(2.12)
$$
will be a recursive operator for eq.(1.1) when
$$
\ti g_n = L_n g_n                                               \eqno(2.13)
$$
is a new generalized symmetry associated to (1.1). This condition
implies that
$$
A(L_n) = L_{n,t} - [f_n^*, L_n] = 0.                            \eqno(2.14)
$$
%--------------------------------------------------------------------
\par
{}From (2.10) it follows that
$$
A(g_n^*) = \pa_{\tau}(f_n^*).                                   \eqno(2.15)
$$
In fact, defining
$$
B_n = \pa_t(g_n) = \sum_k {\pa g_n \over \pa u_{n+k}} f_{n+k},
$$
we have
$$
B_n^* = \sum_m {\pa B_n \over \pa u_{n+m}} D^m =
\sum_{m,k} {\pa^2 g_n \over \pa u_{n+k} \pa u_{n+m}} f_{n+k} D^m +
\sum_{m,k} {\pa g_n \over \pa u_{n+k}} {\pa f_{n+k} \over \pa u_{n+m}} D^m
$$ $$
= \pa_t (g_n^*) + g_n^* f_n^*,                                  \eqno(2.16)
$$
and thus eq.(2.15) is obtained by introducing (2.16) into the Frechet
derivative of (2.10).
%--------------------------------------------------------------------
\par
Eq.(2.15) implies that, as the rhs is an operator of the order 1
(see (2.11)), the highest terms in the lhs must be zero. Thus we can
define an {\it approximate} symmetry of {\it order} $i$ and {\it length}
$m$, i.e. an operator
$$
G_n = \sum_{k=i-m+1}^i g_n^{(k)} D^k
$$
such that the highest $m$ terms of the operator
$$
A(G_n) = \sum_{k=i-m}^{i+1} a_n^{(k)} D^k
$$
are zeroes. Taking into account eq.(2.15), we find that we must have
$i-m+2 > 1$ if the equation
$$
A(G_n) = 0                                                      \eqno(2.17)
$$
is to be true and $G_n$ is to be an approximation to the Frechet
derivative of $g_n$, the local generalized symmetry of our equation.
%--------------------------------------------------------------------
\par
{}From this results we can derive the first integrability condition
which can be stated in the following Theorem:
%--------------------------------------------------------------------
\bigskip
\par
{\sl Theorem 1.} If eq.(1.1) has a generalized local symmetry of order
$i \ge 2$, then it must have a conservation law given by
$$
\pa_t \log f_n^{(1)} = (D-1) q_n^{(1)},                         \eqno(2.18)
$$
where $q_n^{(1)}$ is an RF.
\bigskip
%--------------------------------------------------------------------
\par
The proof of this Theorem is contained in Appendix A.
\par
In this way
we have showed the existence of the first {\it canonical} conservation
law. The next canonical conservation laws could be obtained in the same
way, by assuming a higher symmetry and thus being able to consider
an approximate symmetry of higher length. These canonical conservation
laws would, however, be very complicated (they will depend on the
order of the generalised symmetry) and very difficult to reduce
to simple expressions (not depending on this order). So we prefer to
follow an alternative way which requires the existence of two higher
symmetries.
%--------------------------------------------------------------------
\par
This procedure can be carried out, as we already know one canonical
conservation law. Let us now assume that there exist two RF $g_n$ and
$\ti g_n$ which generate symmetries of order $i$ and $i+1$ respectively.
In correspondence to the symmetries, we can construct the approximate
symmetries $G_n$ and $\ti G_n$ of orders $i$ and $i+1$ respectively,
and from (2.1) $g_n^{(i)}$ and $\ti g_n^{(i+1)}$ will be different
from zero for all $n$ (see (A3), Appendix A). Starting from $G_n$ and
$\ti G_n$, we can construct the operator
$$
\hat G_n = G_n^{-1} \ti G_n.                                    \eqno(2.19)
$$
As from (2.14) we have
$$
A(G_n^{-1}) = - G_n^{-1} A(G_n) G_n^{-1}, \qquad
A(L_n K_n) = A(L_n) K_n + L_n A(K_n),
$$
we obtain
$$
A(\hat G_n) = G_n^{-1} [- A(G_n) \hat G_n + A(\ti G_n)].       \eqno(2.20)
$$
Let's notice that, as $G_n$ is an approximate symmetry, its inverse
will be an operator characterized by an infinite sum. Consequently
$\hat G_n$, thought being  an approximate symmetry of order 1 and
length $i$ (the lowest of the two lengths of $G_n$ and $\ti G_N$) is
represented by an infinite sum. This shows that under the hypothesis
that two local higher simmetries exist, we can reduce ourselves to consider
approximate symmetries of order 1. In such a way $g_n^{(1)} = f_n^{(1)}$
and for $q_n^{(1)}$ the following simple formula can be obtained:
$q_n^{(1)} = g_n^{(0)} - f_n^{(0)}$. We can now state the following
Theorem proved in Appendix B:
%--------------------------------------------------------------------
\bigskip
\par
{\sl Theorem 2.} If the equation (1.1) with the conditions (2.1) has
two generalized local symmetries of order $i$ and $i+1$, with
$i \ge 4$, then the following canonical conservation laws must be true:
$$
\pa_t p_n^{(k)} = (D-1) q_n^{(k)} \qquad (k = 1, 2, 3),
$$ $$
p_n^{(1)} = \log {\pa f_n \over \pa u_{n+1}}, \qquad
p_n^{(2)} = q_n^{(1)} + {\pa f_n \over \pa u_n},
$$ $$
p_n^{(3)} = q_n^{(2)} + {1 \over 2} (p_n^{(2)})^2 +
{\pa f_n \over \pa u_{n+1}} {\pa f_{n+1} \over \pa u_n},        \eqno(2.21)
$$
where $q_n^{(k)}$ ($k = 1, 2, 3$) are some RFs.
\bigskip
%--------------------------------------------------------------------
\par
As we have seen up to now, if eq.(1.1) has local generalised symmetries
of high enough order, we can construct a few canonical conservation
laws depending on the function at the rhs of (1.1).

One can divide the
conservation laws into conjugacy classes under an equivalence condition.
Two conservation laws
$$
p_{n,t} = (D-1) q_n, \qquad r_{n,t} = (D-1) s_n
$$
are said to be {\it equivalent} if
$$
p_n \sim r_n.                                                   \eqno(2.22)
$$
Then we can say that the local conservation law is {\it trivial} if
$p_n \sim 0$. If $p_n \sim r_n(u_n)$, with $r'_n \ne 0$ at least for
some $n$, then we say that the conservation law is of {\it zeroth order}
while if
$$
p_n \sim r_n(u_{n+N}, \ \dots, \ u_n), \qquad N > 0,
$$ and $$
{\pa^2 r_n \over \pa u_n \pa u_{n+N}} \ne 0
$$
for at least some $n$, we say that the conservation law is of
{\it order} $N$.
%--------------------------------------------------------------------
\par
An alternative way to define equivalence classes of local conservation
laws is via the formal variational derivative. If the local conservation
law is trivial, then ${\de p_n \over \de u_n} = 0$, if it is of zeroth
order, then ${\de p_n \over \de u_n} = r'_n(u_n) \ne 0$ for at least some
$n$ while if it is of order $N$, then
$$
{\de p_n \over \de u_n} =
\ti p_n(u_{n+N}, \ \dots, \ u_n, \ \dots, \ u_{n-N}),
$$
where
$$
{\pa \ti p_n \over \pa u_{n+N}} \ne 0, \qquad
{\pa \ti p_n \over \pa u_{n-N}} \ne 0
$$
for at least some $n$.
%--------------------------------------------------------------------
\par
Let us denote by $\ti p_n$ the formal variational derivative of the
density $p_n$ of a local conservation law, i.e.
$$
\ti p_n = {\de p_n \over \de u_n}.                              \eqno(2.23)
$$
For any conserved density $p_n$ we derive
$$
p_{n,t} \sim \ti p_n f_n \sim 0                                 \eqno(2.24)
$$
by direct calculation; then, by carrying out the formal variational derivative
of eq.(2.24), taking into account that in a summation
$$
{\pa \ti p_{n+k} \over \pa u_n} = {\pa \ti p_n \over \pa u_{n+k}},
$$
we get that the formal variational derivative $\ti p_n$ of a conserved
density $p_n$ satisfies the following equation:
$$
(\pa_t + f_n^{*T}) \ti p_n = 0,                                 \eqno(2.25)
$$
where the transposed Frechet derivative of $f_n$ is given by
$$
f_n^{*T} = {\pa f_{n+1} \over \pa u_n} D + {\pa f_n \over \pa u_n}
+ {\pa f_{n-1} \over \pa u_n} D^{-1} =
f_{n+1}^{(-1)} D + f_n^{(0)} + f_{n-1}^{(1)} D^{-1}.            \eqno(2.26)
$$
%--------------------------------------------------------------------
\par
Let us consider the Frechet derivative of $\ti p_n$ for a local
conservation law of order $N$. In such a case, we have
$$
\ti p_n^* = \sum_{k=-N}^N \ti p_n^{(k)} D^k, \qquad
\ti p_n^{(k)} = {\pa \ti p_n \over \pa u_{n+k}}.               \eqno(2.27)
$$
Let us construct the following operator:
$$
B(\ti p_n^*) = \ti p_{n,t}^* + \ti p_n^* f_n^* +
f_n^{*T} \ti p_n^* = \sum_k b_n^{(k)} D^k,                      \eqno(2.28)
$$
where $b_n^{(k)}$ are some RFs. From eq.(2.28) we get that
$$
b_n^{(k)} = \ti p_{n,t}^{(k)} + \sum_j (\ti p_n^{(j)} f_{n+j}^{(k-j)}
+ f_{n+j}^{(-j)} \ti p_{n+j}^{(k-j)}).                          \eqno(2.29)
$$
By differentiating eq.(2.25) with respect to $u_{n+k}$, we can rewrite
eq.(2.29), after a long but straightforward calculation, in the form
$$
b_n^{(k)} = - \sum_j \ti p_{n+j} {\pa^2 f_{n+j}
\over \pa u_n \pa u_{n+k}}                                      \eqno(2.30)
$$
and thus prove that, as $f_n$ depends just on $u_n$ and $u_{n \pm 1}$,
$b_n^{(k)}$ are different from zero only for
$-2 \le k \le 2$.
%--------------------------------------------------------------------
\par
In such a way, for a sufficiently high order conserved density $p_n$,
we can require that
$$
B(\ti p_n^*) = 0                                                \eqno(2.31)
$$
is approximately solved. If the first $m < N-1$ terms of the Frechet
derivative of $\ti p_n$ satisfy eq.(2.31), then we say that we have
an {\it approximate} conserved density of {\it order} $N$ and {\it length}
$m$.

Taking all the results up to now obtained into account, we can state
the following Theorem which will be proved in Appendix C:
%--------------------------------------------------------------------
\bigskip
\par
{\sl Theorem 3.} If the chain (1.1) with conditions (2.1) has a
conservation law of order $N \ge 3$, and the condition (2.18) is satisfied,
then the following relations must take place:
$$
r_n^{(k)} = (D-1) s_n^{(k)} \qquad (k = 1, 2)                   \eqno(2.32a)
$$ with $$
r_n^{(1)} = \log[- f_n^{(1)} / f_n^{(-1)}], \qquad
r_n^{(2)} = s_{n,t}^{(1)} + 2 f_n^{(0)},                        \eqno(2.32b)
$$
where $s_n^{(k)}$ are RFs.
\bigskip
%--------------------------------------------------------------------
\par
We will end this Section by adding a few comments. First of all we can
notice that the request of existence of nontrivial local conservation
laws is more restrictive than that of existence of symmetries. In fact,
there are many instances in which there exist generalized symmetries,
but not nontrivial conservation laws. This is the case, for example,
of the $c$-integrable equations, i.e. such nonlinear equations which
are transformed into a linear equation.
%--------------------------------------------------------------------
\par
If one compares Theorem 1 and Theorem 2, one can think that among
conditions (2.18) and (2.21) with $k = 2, 3$ there is a difference
of importance, as conditions (2.21) require the existence of two
generalized symmetries while for condition (2.18) only one generalized
symmetry is sufficient. However, we could obtain conditions
(2.21) with $k = 2, 3$, assuming that only one symmetry of an order
$i \ge 4$ exists, but calculations in the proof would be much more
difficult.
%--------------------------------------------------------------------
\par
All cases cosidered here required that just $f_n^{(1)} \ne 0$ when
we derived conditions (2.21). An analogous set of conditions could
be derived if we request that just $f_n^{(-1)} \ne 0$ for all $n$.
They can be derived in a straitforward way considering expansions
in negative powers of $D$, instead of positive as we have done here.
This derivation is left to the readers as an exercise. This set of
conditions also will have the form of canonical conservation laws:
$$
\pa_t \hat p_n^{(k)} = (D-1) \hat q_n^{(k)},                    \eqno(2.33)
$$
and conserved densities will be symmetric to ones of (2.21). For
example
$$
\hat p_n^{(1)} = \log {\pa f_n \over \pa u_{n-1}}.              \eqno(2.34)
$$
%--------------------------------------------------------------------
\par
Let us notice moreover that if $H_n^{(1)}$ and $H_n^{(2)}$ are two
solutions of (2.31) of different order, the operator
$$
K_n = (H_n^{(1)})^{-1} H_n^{(2)}                                \eqno(2.35)
$$
satisfies (2.14) and thus it is a recursive operator. Consequently,
if we start from two approximate solutions of (2.31), i.e. two
Frechet derivatives of formal variational derivatives of conserved
densities, we can, using (2.31), get an approximate symmetry.
So, one can derive all the conditions (2.21), (2.32), (2.33),
assuming the existence of two higher order local conservation laws.
In the case of conditions (2.33), one should use the same formula
(2.35), but $(H_n^{(1)})^{-1}$ will be a series in positive powers
of the shift operator $D$:
$$
(H_n^{(1)})^{-1} = \sum_{k=-N}^{+\infty} h_n^{(k)} D^k.
$$
This proves the statement written before that conservation laws are
"more fundamental" than symmetries.
%--------------------------------------------------------------------
\par
If we compare conditions (2.21), (2.32), (2.33), we can see that,
for example,
$$
r_n^{(1)} = \log(-1) + p_n^{(1)} - \hat p_n^{(1)},
$$
i.e. the first of conditions (2.32) means that
$p_n^{(1)} \sim \hat p_n^{(1)}$, i.e. first canonical conservation laws
of (2.21) and (2.33) are equivalent. The same result could be obtained for the
second condition of (2.32). In particular, the set of conditions (2.33) can
be derived starting from conditions (2.21) and (2.32). However, these
conditions are of great importance in themselves, as there might be
equations of interest which satisfy (2.21) and (2.33), but not (2.32).
%--------------------------------------------------------------------
\par
The solution of the conditions, be those obtained by requesting existence
of the generalised symmetries or those of local conservation laws,
provide the highest order coefficients of the Frechet derivative
of a symmetry or of the formal variational derivative of a
conserved density. Those coefficients are the building blocks for the
reconstruction of the symmetries or of the formal variational derivatives
of the conserved densities. In fact, limiting ourselves to the case of
symmetries for the sake of concreteness,
$g_n^{(k)} = {\pa g_n \over \pa u_{n+k}}$ for $k = i, i-1, \dots$.
Thus the knowledge $g_n^{(k)}$ for a few values of $k$ gives a set of
partial differential equations for $g_n$ in terms of its variable
whose solution provides the needed symmetry. The same way allows us
to reconstruct variational derivatives of conserved densities.
There is, however, a more direct way to obtain conserved densities.
In fact, if we know highest coefficients of the first order solution $L_n$ of
eq.(2.14), we can obtain several conserved densities by the following
formula:
$$
p_n^{(j)} = \res(L_n^j) \qquad (j = 1, \ 2, \ \dots)
$$
(see Appendix B). Moreover, the solution of the conditions (2.21)
leads to three local conservation laws in explicit form.
%%-------------------------------------- Section 3 ------------------
\bigskip
\medskip
\par
{\bf 3. Introduction to chain classification}
\medskip
\par
In this Section, using just two conditions, we give two Theorems which
split eq.(1.1) into two classes of nonlinear equations
depending on a limited number of variables. The complete classification
using all other conditions will be presented in the second part of this
work. Here, however, we present at the end examples of nonlinear
evolution equations on the lattice belonging to each of the obtained classes.
%%-------------------------------------- Section 3.1 ----------------
\bigskip
\medskip
\par
{\sl 3.1 Main classification Theorems}
\medskip
\par
Let us apply the first of the conditions (2.32) to eq. (1.1), (2.1),
i.e.
$$
r_n^{(1)} = \log[- f_n^{(1)} / f_n^{(-1)}] = (D-1) s_n^{(1)}.   \eqno(3.1)
$$
In such a case,
$$
r_n^{(1)} = r_n^{(1)}(u_{n+1}, u_n, u_{n-1})
$$
and consequently $s_n^{(1)} = s_n^{(1)}(u_n, u_{n-1})$ is defined
uniquely if $s_n^{(1)}$ is to be a RF.
%--------------------------------------------------------------------
\par
Introducing the function
$$
y_n = y_n(u_{n+1}, u_n) = \exp s_{n+1}^{(1)},
$$
we get from (3.1) that $y_n$ must satisfy the following difference
equation:
$$
y_n f_n^{(-1)} + y_{n-1} f_n^{(1)} = 0                          \eqno(3.2)
$$
where the function $y_n$ is, by definition, in a generic point of the space
of its independent variables, always different from zero. Eq.(3.2)
is a constraining equation for $f_n$ which thus provide a reduction of
$f_n$ in terms of new symmetry variables. In fact, defining a function
$v_n(u_{n+1}, u_n, u_{n-1})$ so that
$$
y_n = {\pa v_n \over \pa u_{n+1}}, \qquad
y_{n-1} = - {\pa v_n \over \pa u_{n-1}},                        \eqno(3.3)
$$
we find that (3.2) is identically satisfied if
$f_n(u_{n+1}, u_n, u_{n-1})$ is written as
$$
f_n = \ph_n(u_n, v_n).                                          \eqno(3.4)
$$
Then, as
$$
f_n^{(1)} = {\pa \ph_n \over \pa v_n} y_n, \qquad
f_n^{(-1)} = - {\pa \ph_n \over \pa v_n} y_{n-1},
$$
we get that
$$
{\pa \ph_n \over \pa v_n} \ne 0 \quad \forall n.                \eqno(3.5)
$$
%--------------------------------------------------------------------
\par
Eq.(3.3) can be "solved" by introducing a function
$z_n = z_n(u_{n+1}, u_n)$ such that
$$
y_n = {\pa^2 z_n \over \pa u_{n+1} \pa u_n}                     \eqno(3.6)
$$ and $$
v_n = {\pa \over \pa u_n}(z_n - z_{n-1}).                       \eqno(3.7)
$$
We can now state the following Theorem:
%--------------------------------------------------------------------
\bigskip
\par
{\sl Theorem 1.} The condition (3.1) is satisfied iff the rhs of (1.1),
taking into account conditions (2.1), can be represented as a function
(3.4) with $v_n$ given by (3.7) in term of a function
$z_n(u_{n+1}, u_n)$ such that
$$
{\pa^2 z_n \over \pa u_{n+1} \pa u_n} \ne 0 \quad \forall n.    \eqno(3.8)
$$
The function $\ph_n$ will depend in an essential way from the
variable $v_n$ (see (3.5)).
\bigskip
%--------------------------------------------------------------------
\par
The proof is already contained in the results presented before; (3.8)
is given by (3.6), as $y_n$ is always different from zero. In such a
way, we have reduced the problem of finding a set of functions of three
variables to that of finding two sets of functions of two variables only.
Even if $v_n$ depends on $u_n$, the variables $v_n$ and $u_n$ may be
taken to be independent variables, as $y_n$ is always different from
zero.
%--------------------------------------------------------------------
\par
We can now pass to consider the first of the canonical conservation laws
given by (2.21), i.e.
$$
p_{n,t}^{(1)} \sim 0, \qquad p_n^{(1)} = \log f_n^{(1)} =
\log \left( {\pa \ph_n \over \pa v_n} y_n \right).              \eqno(3.9)
$$
Let us explicitate (3.9); we have
$$
\Ph_n = {\pa p_n^{(1)} \over \pa u_{n+1}} f_{n+1} +
{\pa p_n^{(1)} \over \pa u_n} f_n +
{\pa p_{n+1}^{(1)} \over \pa u_n} f_n \sim
p_{n,t}^{(1)} \sim 0,                                           \eqno(3.10)
$$ $$
\Ph_n = \Ph_n(u_{n+2}, \ u_{n+1}, \ u_n, \ u_{n-1}).
$$
As $\Ph_n$ is a RF equivalent to zero, we have
$$
{\pa^2 \Ph_n \over \pa u_{n+2} \pa u_{n-1}} =
{\pa^2 p_n^{(1)} \over \pa u_{n+1} \pa u_{n-1}}
{\pa f_{n+1} \over \pa u_{n+2}} +
{\pa^2 p_{n+1}^{(1)} \over \pa u_{n+2} \pa u_n}
{\pa f_n \over \pa u_{n-1}} = 0 \quad \forall n.                \eqno(3.11)
$$
{}From the definition of $p_n^{(1)}$ in terms of $\ph_n$ and $y_n$,
we get
$$
{\pa^2 \over \pa v_{n+1}^2}
\left( \log {\pa \ph_{n+1} \over \pa v_{n+1}} \right) /
{\pa \ph_{n+1} \over \pa v_{n+1}} =
{\pa^2 \over \pa v_n^2}
\left( \log {\pa \ph_n \over \pa v_n} \right) /
{\pa \ph_n \over \pa v_n} \quad \forall n.                      \eqno(3.12)
$$
To obtain (3.12), we have used the fact that $y_n$ and
${\pa \ph_n \over \pa v_n}$ are both different from zero for any $n$.
Eq.(3.12) implies that the function
$\ph_n(u_n, v_n)$ must satisfy the following ordinary differential
equation:
$$
{\pa \ph_n \over \pa v_n} =
\ep \ph_n^2 + a_n(u_n) \ph_n + b_n(u_n),                        \eqno(3.13)
$$
where $\ep$ is an arbitrary $n$-independent constant, and $a_n$ and
$b_n$ are two arbitrary $u_n$-depen\-dent sets of functions.
\bigskip
%--------------------------------------------------------------------
\par
Eq.(3.13) implies that we have two main different cases, characterised
by the form of the dependence of $\ph_n$ on $v_n$: the case $\ep \ne 0$
and the case $\ep = 0$. If $\ep \ne 0$ and, for example,
$$
\De_n(u_n) = a_n^2 - 4 \ep b_n \ne 0 \qquad \forall n,          \eqno(3.14)
$$
then
$$
\ph_n = {\De_n^{1/2} \over 2 \ep}
\tanh \left( {\De_n^{1/2} \over 2} (K_n(u_n) - v_n) \right)
- {a_n \over 2 \ep},                                            \eqno(3.15)
$$
where $K_n$ is an arbitrary integration $u_n$-dependent set of functions.
If
$$
\ep \ne 0, \qquad \De_n = 0 \quad \forall n,
$$
we are led to
$$
\ph_n =  {1 \over \ep (K_n(u_n) - v_n)}
- {a_n \over 2 \ep}.                                            \eqno(3.16)
$$
In general, if we consider the case $\ep \ne 0$, the function $\ph_n$
may have (at least theoretically) the form (3.15) for some $n$ and
the form (3.16) for some other $n$. At the end of this Section we will
write down an integrable example of this kind.
%--------------------------------------------------------------------
\par
Let us consider the case $\ep = 0$. Eq.(3.13) is equivalent to the
following ordinary differential equation:
$$
{\pa \ph_n \over \pa v_n} =
\exp[a_n(u_n) v_n + c_n(u_n)],                                  \eqno(3.17)
$$
where $c_n$ is an arbitrary $u_n$-dependent set of functions. It is easy
to show that the function $\ph_n$ must have the form:
$$
\ph_n = \al_n(u_n) \exp(a_n v_n) + \be_n(u_n) v_n + \ga_n(u_n).  \eqno(3.18)
$$
Using (3.13) or the equivalent equation (3.17), we can obtain conditions
for coefficients $\al_n$, $\be_n$, $\ga_n$. In particular,
$\al_n \be_n = 0$ for any $n$. Below readers can see an example such that
$$
\al_n = \ti \al_n(u_n) P_n, \qquad  \be_n = \ti \be_n(u_n) Q_n,
$$
where $P_n$ and $Q_n$ are two orthogonal projection operators defined by:
$$
P_n = {1-(-1)^n \over 2}, \qquad Q_n = {1+(-1)^n \over 2}.      \eqno(3.19)
$$
It is clear that the equation $\al_n \be_n = 0$ has not only such the
solution; the Toda and Volterra equations correspond to the solution:
$\al_n = 0$, $\be_n \ne 0$ for any $n$.
%--------------------------------------------------------------------
\par
Eqs. (3.15), (3.16), (3.18) provide the two
classes of nonlinear differential-difference equations of type (1.1)
which satisfy the simplest integrability condition of (2.32) and
partially the condition (2.21) with $k=1$. The other conditions
(2.21) and (2.32) with respect to an arbitrary point transformation
will reduce almost completely the classes of equations to a few
equations only.
%%-------------------------------------- Section 3.2 ----------------
\bigskip
\medskip
\par
{\sl 3.2 Examples}
\medskip
\par
Let us consider an example of equation belonging to the first class,
more precisely to the class
$$
\ep \ne 0, \qquad \De_n \ne 0 \quad \forall n.
$$
In such a case we have to fix the arbitrary functions $K_n(u_n)$,
$a_n(u_n)$ and $b_n(u_n)$, the arbitrary constant $\ep$, and to define
the function $v_n$ as function of $u_{n+1}$, $u_n$, $u_{n-1}$ to get
a well defined nonlinear evolution chain equation.
%--------------------------------------------------------------------
\par
Let us set:
$$
K_n = 0, \qquad a_n = 0, \qquad b_n = {1 \over 2}
[p(u_n) r(u_n) - q^2(u_n)], \qquad \ep = {1 \over 2},           \eqno(3.20)
$$
where
$$
p(u_n) = \al u_n^2 + 2 \be u_n + \ga,                           \eqno(3.21a)
$$ $$
r(u_n) = \ga u_n^2 + 2 \de u_n + \om,                           \eqno(3.21b)
$$ $$
q(u_n) = \be u_n^2 + \la u_n + \de.                             \eqno(3.21c)
$$
Then we can define
$$
v_n = - \ {2 \over \sqrt{- 2 b_n}} \ \hbox{arctanh}
\left( {p(u_n) u_{n+1} u_{n-1} + q(u_n) (u_{n+1} + u_{n-1}) + r(u_n)
\over \sqrt{- 2 b_n} \ (u_{n+1} - u_{n-1})} \right)             \eqno(3.22)
$$
which corresponds to define
$$
y_n = {\mu \over p(u_n) u_{n+1}^2 + 2 q(u_n) u_{n+1} + r(u_n)}.
$$
Inserting (3.20), (3.22) into (3.15), we get the following evolutionary
nonlinear chain:
$$
u_{n,t} = {p(u_n) u_{n+1} u_{n-1} + q(u_n) (u_{n+1} + u_{n-1}) + r(u_n)
\over u_{n+1} - u_{n-1}}.                                       \eqno(3.23)
$$
%--------------------------------------------------------------------
\par
Eq.(3.23) depends on 6 arbitrary complex constants (see (3.21)) and is
invariant under linear-fractional transformations (only coefficients
of the polinomials $p$, $q$, $r$ are changed). It was obtained for
the first time in [2] when classifying discrete evolutionary
equations of the form:
$$
u_{n,t} = f(u_{n+1}, \ u_n, \ u_{n-1}).                         \eqno(3.24)
$$
It satisfies all the five integrability conditions, has an infinite set
of higher local conservation laws and should have an infinite set of
generalized symmetries (but nobody has yet proved it). It is the only
example of nonlinear chain of the form (3.24) which cannot be reduced
to the Toda or Volterra equations by Miura transformations. By carrying
out the continuous limit, as one does for the Volterra equation to
obtain the Korteweg-de Vries equation, we get the well-known
Krichever-Novikov equation [8]:
$$
u_t = u_{xxx} - {3 \over 2} {u_{xx}^2 \over u_x} +
{R(u) \over u_x},                                               \eqno(3.25)
$$
where $R(u)$ is an arbitrary 4th degree polinomial of its argument
with constant coefficients.
%--------------------------------------------------------------------
\par
In the second case $\ep = 0$, let us restrict ourselves to the subcase
$\al_n = 0$ for any $n$ (see (3.18)) and consider functions $\ph_n$ of
the form:
$$
\ph_n = \be_n(u_n) v_n + \ga_n(u_n).                            \eqno(3.26)
$$
Let
$$
\ga_n = 0, \qquad v_n = u_{n+1} - u_{n-1},                      \eqno(3.27)
$$
but $\be_n$ be an arbitrary nonzero (for all $n$) function of its
argument. In such a case,
$$
f_n = \be_n(u_n) (u_{n+1} - u_{n-1}).                           \eqno(3.28)
$$
%--------------------------------------------------------------------
\par
The first canonical conservation law implies that
$$
\pa_t \log \be_n \sim 0,
$$ i.e. $$
\be'_n (u_{n+1} - u_{n-1}) \sim \be'_n u_{n+1} -
\be'_{n+1} u_n \sim 0                                           \eqno(3.29)
$$
from which we derive that $\be''_n$ must be an $n$-independent constant:
$$
\be_n = A u_n^2 + B_n u_n + C_n.                                \eqno(3.30)
$$
Inserting this result into (3.29), we get that
$$
(B_{n-1} - B_{n+1}) u_n \sim 0, \qquad \hbox{i.e.}
\qquad B_n = B + (-1)^n \ti B.
$$
Consequently, the first canonical conservation law reads:
$$
\pa_t \log \be_n = (D-1) q_n^{(1)},
$$ where $$
q_n^{(1)} = 2 A u_n u_{n-1} + B (u_n + u_{n-1}) -
\ti B [(-1)^n u_n + (-1)^{n-1} u_{n-1}].                        \eqno(3.31)
$$
%--------------------------------------------------------------------
\par
Introducing (3.28) and (3.31) into the second canonical conservation
law, we get, after a straightforward but lengthty calculation, that
$$
p_{n,t}^{(2)} \sim (C_{n-1} - C_{n+1}) (A u_n^2 + B_n u_n) \sim 0.
$$
We can see, in particular, that if $A \ne 0$, then
$$
C_n = C + (-1)^n \ti C.                                         \eqno(3.32)
$$
Using the last canonical conservation law, one can prove that $C_n$
always must have the form (3.32). So, we are led to the chain
$$
u_{n,t} = [A u_n^2 + (B + (-1)^n \ti B) u_n
+ C + (-1)^n \ti C] (u_{n+1} - u_{n-1})                         \eqno(3.33)
$$
with 5 arbitrary complex constants.
%--------------------------------------------------------------------
\par
By direct calculations, one can then prove that this nonlinear
equation (3.33), for any choice of the arbitrary constants, satisfies
all three canonical conservation laws (2.21) and both additional
conditions (2.32). By obvious point transformations, we can reduce
any nonlinear chain of the form (3.33) to one of the following chains:
$$
u_{n,t} = \be_n(u_n) (u_{n+1} - u_{n-1}),                       \eqno(3.34)
$$ where $$
\be_n = P_n u_n + Q_n                                           \eqno(3.35)
$$
(the Toda chain),
$$
\be_n = u_n                                                     \eqno(3.36)
$$
(the Volterra equation) or
$$
\be_n = C_n - u_n^2,                                            \eqno(3.37)
$$
and in the last case,
$$
C_n = 1, \qquad C_n = 0 \qquad \hbox{or} \qquad C_n = P_n       \eqno(3.38)
$$
(three modifications of the Volterra equation). Here $P_n$ and $Q_n$ are
the orthogonal projection operators (3.19). Unlike the discrete version
of the Krichever-Novikov equation (3.23), (3.21), all the chains (3.33)
can be reduced to the Toda chain by Miura transformations. For example,
in the case of the Volterra equation, we have the transformation
$$
\ti u_n = P_n u_{n+1} u_n + Q_n (u_{n+1} + u_n).                \eqno(3.39)
$$
Formula (3.39) brings any solution $u_n$ of the Volterra equation into
a solution $\ti u_n$ of the Toda chain. Transformations of the modified
Volterra equations (3.34), (3.37), (3.38) into the Volterra equation
are given by the formula
$$
\ti u_n = (C_n + u_n) (C_{n+1} - u_{n+1})                       \eqno(3.40)
$$
in all the three cases. Transformations (3.39), (3.40) together with
point transformations enable one to prove that any nonlinear chain of the
form (3.33) possesses local conservation laws of an arbitrary high order.
This means, in particular, that the chains (3.33) satisfy not only the
classifying conditions (2.21), (2.32) but also all conditions we could
derive using approximate symmetries and conserved densities.
%--------------------------------------------------------------------
\par
The last class of equations we discuss here is given by:
$$
u_{n,t} = F_n(u_{n+1} - u_{n-1}), \qquad
F'_n = A F_n^2 + B_n F_n + C_n,                                 \eqno(3.41)
$$ where $$
B_n = B + (-1)^n \ti B, \qquad C_n = C + (-1)^n \ti C,
$$
and $A$, $B$, $\ti B$, $C$, $\ti C$ are arbitrary complex constants.
These equations have infinite sets of local conservation laws and
satisfy all the five classifying conditions because are reduced to
chains of the form (3.33) by the following transformation:
$$
\ti u_n = F_n(u_{n+1} - u_{n-1}).
$$
In the case
$$
F_n(z) = P_n \tanh z + Q_n z^{-1},                             \eqno(3.42)
$$
one has the differential equation
$$
F'_n = P_n - F_n^2,
$$
i.e. function (3.42) defines an equation of the form (3.41) corresponding
to the last of the modified Volterra equations (see (3.34), (3.37), (3.38)).
So, we have obtained an example of the chain with the rhs which can be
represented as function $\ph_n(u_n, v_n)$  (here $v_n = u_{n+1} - u_{n-1}$)
being $\tanh v_n$ for odd $n$ and $v_n^{-1}$ for even $n$.
The second example of this kind is given by:
$$
F_n(z) = P_n \exp z + Q_n z.
$$
In this case,
$$
F'_n = P_n F_n + Q_n,
$$
i.e. the equation of the form (3.41) is reduced to the Toda chain.
%%-------------------------------------- Appendix A -----------------
\bigskip
\medskip
\par
{\bf Appendix A:} {\sl Proof of Theorem 1}
\medskip
\par
For a sufficiently high order symmetry, i.e. $i >> 1$, the highest
terms of
$$
g_n^* = \sum_{k=j}^i g_n^{(k)} D^k
$$
will satisfy the following equation:
$$
\sum_{l=2}^i g_{n,t}^{(l)} D^l +
\sum_k^i \sum_{m=-1}^1 g_n^{(k)} f_{n+k}^{(m)} D^{k+m} -
\sum_k^i \sum_{m=-1}^1 f_n^{(m)} g_{n+m}^{(k)} D^{k+m} = 0,     \eqno(A1)
$$
where the sum is over those $k$ such that $k + m > 1$, as otherwise
the lhs of (A1) is no more zero. In (A1) the coefficients of any power
of $D$ must vanish; so the coefficients of $D^{i+1}$ reads:
$$
g_n^{(i)} f_{n+i}^{(1)} = f_n^{(1)} g_{n+1}^{(i)}.              \eqno(A2)
$$
As, due to (2.1), $f_n^{(1)} \ne 0$ $\forall n$, we have
$$
g_n^{(i)} = \prod_{k=n}^{n+i-1} f_k^{(1)},                      \eqno(A3)
$$
where we have, with no rectriction, set to unit the "integration" constant.
%--------------------------------------------------------------------
\par
Let us consider now the coefficient of $D^i$; this comes from more
than one term ($k=i$, $m=0$ or $k=i-1$, $m=1$) and involves the time
evolution of $g_n^{(i)}$. It can be cast in the following form:
$$
{g_{n,t}^{(i)} \over g_n^{(i)}} =
{g_{n+1}^{(i-1)} \over \prod_{k=n+1}^{n+i-1} f_k^{(1)}} -
{g_n^{(i-1)} \over \prod_{k=n}^{n+i-2} f_k^{(1)}} -
(D-1) \sum_{k=n}^{n+i-1} f_k^{(0)}                              \eqno(A4)
$$
from which we derive:
$$
\pa_t \log g_n^{(i)} = (D-1) \left[
{g_n^{(i-1)} \over \prod_{k=n}^{n+i-2} f_k^{(1)}} -
\sum_{k=n}^{n+i-1} f_k^{(0)}  \right].                          \eqno(A5)
$$
Introducing (A3) into the lhs of (A5), we can write it as:
$$
\pa_t \log \prod_{k=n}^{n+i-1} f_k^{(1)} =
\sum_{k=n}^{n+i-1} \pa_t \log f_k^{(1)} \sim
i \ \pa_t \log f_n^{(1)} \sim 0 \qquad \hbox{c.v.d.}
$$
%%-------------------------------------- Appendix B -----------------
\bigskip
\medskip
\par
{\bf Appendix B:} {\sl Proof of Theorem 2}
\medskip
\par
As a consequence of Theorem 1, we can assert that we have an approximate
symmetry of order $i=1$
$$
G_n = g_n^{(1)} D + g_n^{(0)} + g_n^{(-1)} D^{-1} +
g_n^{(-2)} D^{-2} + \dots,                                      \eqno(B1)
$$
where from (A3)
$$
g_n^{(1)} = f_n^{(1)}.                                          \eqno(B2)
$$
Instead of (A5) we have
$$
\pa_t \log f_n^{(1)} = (D-1) (g_n^{(0)} - f_n^{(0)})            \eqno(B3)
$$
(see the coefficient of $D$ in the equation $A(G_n) = 0$ with $A$
defined by (2.14)). Consequently, the function of the rhs of the first
canonical conservation law is:
$$
q_n^{(1)} = g_n^{(0)} - f_n^{(0)}
$$
from which we derive
$$
g_n^{(0)} = f_n^{(0)} + q_n^{(1)}.                              \eqno(B4)
$$
%--------------------------------------------------------------------
\par
Let us now consider the coefficient of $D^0$ in the equation $A(G_n) = 0$:
$$
\pa_t g_n^{(0)} = (D-1) [f_{n-1}^{(1)}
(g_n^{(-1)} - f_n^{(-1)})].                                     \eqno(B5)
$$
Eq.(B5) is the second canonical conservation law with the lhs given
by (B4):
$$
p_n^{(2)} = g_n^{(0)} = f_n^{(0)} + q_n^{(1)}, \qquad
q_n^{(2)} = f_{n-1}^{(1)} (g_n^{(-1)} - f_n^{(-1)}).            \eqno(B6)
$$
{}From (B6) we get
$$
g_n^{(-1)} = f_n^{(-1)} + q_n^{(2)} / f_{n-1}^{(1)}.            \eqno(B7)
$$
\par
To get the last relation, it is simpler to use the following Lemma:
%--------------------------------------------------------------------
\bigskip
\par
{\sl Lemma.} If $ H_n = h_n^{(i)} D^i + h_n^{(i-1)} D^{i-1} + \dots $ is
an approximate symmetry, which satisfies the first $m \ge i+2$ terms
of the equation
$H_{n,t} = [f_n^*, H_n]$, then
$$
\res(H_n) \equiv h_n^{(0)}                                      \eqno(B8)
$$
will be a conserved density.
\bigskip
%--------------------------------------------------------------------
\par
In fact,
$$
\pa_t \res(H_n) = \res(H_{n,t}) = \res [f_n^*, H_n].            \eqno(B9)
$$
The coefficient of $D^0$ of $[f_n^*, H_n]$ can be obtained only from
terms of the type
$$
[r_n D^m, s_n D^{-m}]
$$
which are equivalent to zero.
%--------------------------------------------------------------------
\par
As any power of an approximate symmetry is also an approximate symmetry
of the same lengh,
we can construct a new conserved density, calculating the residue of
$G_n^2$. In such a case, after a long but straightforward calculation,
we get
$$
\res(G_n^2) = \res[(g_n^{(1)} D + g_n^{(0)} + g_n^{(-1)} D^{-1} +
\dots)^2] =
$$ $$
g_n^{(1)} g_{n+1}^{(-1)} + (g_n^{(0)})^2 + g_n^{(-1)} g_{n-1}^{(1)}
\sim 2 g_n^{(1)} g_{n+1}^{(-1)} + (g_n^{(0)})^2 \sim
$$ $$
2 q_n^{(2)} + (p_n^{(2)})^2 + 2 f_n^{(1)} f_{n+1}^{(-1)}
= 2 p_n^{(3)}.
$$
As, from the previous Lemma, if $\pa_t \res G_n^2 \sim 0$, so will be
$\pa_t p_n^{(3)} \sim 0$ c.v.d.
%%-------------------------------------- Appendix C -----------------
\bigskip
\medskip
\par
{\bf Appendix C:} {\sl Proof of Theorem 3}
\medskip
\par
Let us assume that we have a solution of
$$
\pa_t P_n + P_n f_n^* + f_n^{*T} P_n = 0,                       \eqno(C1)
$$
where $P_n$ is an approximate conserved density
$$
P_n = \sum_{k=N-m+1}^N p_n^{(k)} D^k                            \eqno(C2)
$$
of order $N \ge 3$ and length $m \ge 2$. In such a case, introducing
(C2) into (C1), the coefficient of $D^{N+1}$ reads:
$$
p_n^{(N)} f_{n+N}^{(1)} + f_{n+1}^{(-1)} p_{n+1}^{(N)} = 0.     \eqno(C3)
$$
As $f_n^{(\pm 1)} \ne 0$ for any $n$, we can prove that
$p_n^{(N)} \ne 0$ for any $n$. Then we get
$$
- f_{n+N}^{(1)} / f_{n+1}^{(-1)} = p_{n+1}^{(N)} / p_n^{(N)}    \eqno(C4)
$$
and thus, by taking the logarithm of both sides, we are led to
$$
r_n^{(1)} = \log[- f_n^{(1)} / f_n^{(-1)}].                     \eqno(C5)
$$
%--------------------------------------------------------------------
\par
{}From (C4) and (C5) we get
$$
p_n^{(N)} = s_{n+1} f_{n+1}^{(1)} f_{n+2}^{(1)}
\dots f_{n+N-1}^{(1)},                                          \eqno(C6)
$$
where $s_n$ is such that
$$
s_n f_n^{(1)} + f_n^{(-1)} s_{n+1} = 0.                         \eqno(C7)
$$
The coefficient at order $D^N$ of (C1) gives:
$$
\pa_t \log(s_{n+1} f_{n+1}^{(1)} \dots f_{n+N-1}^{(1)}) +
f_n^{(0)} + f_{n+N}^{(0)} \sim 0
$$
from which
$$
(N-1) \ \pa_t \log f_n^{(1)} + \pa_t \log s_n + 2 f_n^{(0)} \sim 0.
$$
As condition (2.18) is satisfied, and $\log s_n = s_n^{(1)}$
(compare (C7) and the first of the conditions (2.32)), we are led to
$$
r_n^{(2)} = \pa_t s_n^{(1)} + 2 f_n^{(0)} \qquad \hbox{c.v.d.}
$$
%%--------------------------------------Acknowledgments--------------
\bigskip
\medskip
\par
{\bf Acknowledgments}
\par
\medskip
The research of R.Y. is partially supported by grants from INTAS  and
Russian Foundation for Fundamental Research. The visits to Rome of R.Y.
had been possible thanks to a fellowshop from GNFM of CNR.

%%-------------------------------------- References -----------------
\bigskip
\medskip
\par
{\bf References}
\par
\medskip
{\bf 1.} M.J. Ablowitz and J. Ladik, {\it Nonlinear differential-difference
equations}, J. Math. \break Phys. 16 (1975) 598-603; {\it Nonlinear
differential-difference equations and Fourier transform}, J. Math. Phys.
17 (1976) 1011-1018; {\it On the solution of a class of nonlinear partial
difference equations}, Stud. Appl. Math. 55 (1976) 213; {\it On the solution
of a class of nonlinear partial difference equations}, Stud. Appl. Math. 57
(1976) 1.
\par
\medskip
{\bf 2.} R.I. Yamilov, {\it Classification of discrete evolution
equations}, Uspekhi Mat. Nauk 38:6 (1983) 155-156 [in  Russian].
\par
\medskip
{\bf 3.} R.I. Yamilov, {\it Generalizations of the Toda model,
and conservation laws}, Preprint, Institute of Mathematics, Ufa,
1989 [in Russian];
\par
English version: R.I. Yamilov, {\it Classification of Toda type
scalar lattices}, Proc. Int. Conf. NEEDS'92,
World Scientific Publishing, 1993, p. 423-431.
\par
\medskip
{\bf 4.} A.B. Shabat and R.I. Yamilov, {\it Symmetries of nonlinear
chains}, Algebra i Analiz 2:2 (1990) 183-208 [in Russian];
English transl. in Leningrad Math. J. 2:2 (1991) 377-400.
\par
\medskip
{\bf 5.} V.V. Sokolov and A.B. Shabat, {\it Classification of integrable
evolution equations}, Soviet Scientific Rev., Section C, Math. Phys. Rev.
4 (1984) 221-280.
\par
\medskip
{\bf 6.} A.V. Mikhailov, A.B. Shabat, and R.I. Yamilov,
{\it The symmetry approach to the classification of nonlinear equations.
Complete lists of integrable systems},
Uspekhi Mat. Nauk 42:4 (1987) 3-53 [in Russian];
English transl. in Russian Math. Surveys 42:4 (1987) 1-63.
\par
\medskip
{\bf 7.} A.V. Mikhailov, A.B. Shabat, and V.V. Sokolov,
{\it The symmetry approach to classification of integrable equations},
in "What is Integrability?", Springer-Verlag (Springer Series in
Nonlinear Dynamics), 1991, p. 115-184.
\par
\medskip
{\bf 8.} I.M. Krichever and S.P. Novikov, {\it Holomorphic bundles
over algebraic curves and non-linear equations},
Uspekhi Mat. Nauk 35 (1980) 47-68 [in Russian];
English transl. in Russian Math. Surveys 35 (1980) 53-80.
%--------------------------------------------------------------------
\end